\documentclass[11pt]{article}
\usepackage{amsfonts,epsfig,graphicx,subfigure,url,wrapfig}

\title{Amazons is PSPACE-complete}
 \author{Robert A. Hearn%
         \thanks{MIT Computer Science and Artificial Intelligence Laboratory,
                32 Vassar Street, Cambridge, MA 02139, U.S.A.,
                \protect\url{rah@ai.mit.edu}}
 }
\date{}

\newcommand{\fig}[1]{Figure~\ref{fig:#1}}
\newcommand{\figs}{Figures}

\newtheorem{thm}{Theorem}
\newtheorem{lem}[thm]{Lemma}

\newcommand{\pf}{\noindent{\em Proof.}\hspace{1em}}

\makeatletter
\def\GrabProofArgument[#1]{ (#1): \egroup\ignorespaces}

\def\GrabProofArgument[#1]{ #1: \egroup\ignorespaces}
\def\pf{\noindent\textbf\bgroup Proof%
           \@ifnextchar[{\GrabProofArgument}{: \egroup\ignorespaces}}

\makeatother

\advance \topmargin by -\headheight
\advance \topmargin by -\headsep
\evensidemargin \oddsidemargin
\let\latexcite=\cite
\def\cite{\nolinebreak\latexcite}
\let\latexref=\ref
\def\ref{\nolinebreak\latexref}

{\makeatletter
 \gdef\xxxmark{%
   \expandafter\ifx\csname @captype\endcsname\relax % based on a TeXBook e.g.
     \marginpar{xxx}% not in a caption, can use marginpar
   \else
     xxx % notice trailing space
   \fi}
 \gdef\xxx{\@ifnextchar[\xxx@lab\xxx@nolab}
 \long\gdef\xxx@lab[#1]#2{{\bf [\xxxmark #2 ---{\sc #1}]}}
 \long\gdef\xxx@nolab#1{{\bf [\xxxmark #1]}}
\long\gdef\xxx@lab[#1]#2{}\long\gdef\xxx@nolab#1{}%
}

{\makeatletter \gdef\fps@figure{!htbp}}

\newenvironment{proofsketch}{\begin{proof}[sketch]}{\end{proof}}
\let\realendproofsketch=\endproofsketch
\def\endproofsketch{\hspace*{\fill}$\Box$\realendproofsketch}

\begin{document}

\maketitle

\begin{abstract}
Amazons is a board game which combines elements of Chess and Go. It has become popular in recent years, and has served as a useful platform for both game-theoretic study and AI games research.
Buro \cite{Buro-2000} showed that simple Amazons endgames are NP-equivalent, leaving the complexity of the general case as an open problem. 

We settle this problem, by showing that deciding the outcome of an $n\times n$ Amazons position is PSPACE-hard. We give a reduction from one of the PSPACE-complete two-player formula games described by Schaefer \cite{Schaefer-1978-games}. Since the number of moves in an Amazons game is polynomially bounded (unlike Chess and Go), Amazons is in PSPACE. It is thus on a par with other two-player, bounded-move, perfect-information games such as Hex \cite{Even-Tarjan-1976, Reisch-1981}, Othello \cite{Iwata-1994}, and Kayles \cite{Schaefer-1978-games}. Our construction also provides an alternate proof that simple Amazons endgames are NP-equivalent.

Our reduction uses a number of amazons polynomial in the input formula length; a remaining open problem is the complexity of Amazons when only a constant number of amazons is used.

\end{abstract}

\section{Introduction}

\paragraph{Background.}
Amazons was invented by Walter Zamkauskas in 1988. Both human and computer opponents are available for Internet play, and there have been several tournaments, both for humans and for computers.

Amazons has several properties which make it interesting for theoretical study.
Like Go, its endgames naturally separate into independent subgames; these have been studied using combinatorial game theory \cite{Berlekamp-2001, Snatzke-2004}.
%Though played on a small board, 
Amazons has a very large number of moves available from a typical position, even more than in Go. This makes straightforward search algorithms impractical for computer play. As a result, computer programs need to incorporate more high-level knowledge of Amazons strategy \cite{Muller-Tegos-2002, Lieberum-2005}. By showing that generalized Amazons is PSPACE-complete, we provide strong evidence that there is a practical limit to the degree of analysis possible from an arbitrary position.

\paragraph{Amazons Rules.}

\begin{figure}
\centering
\hspace{.6 in}
\subfigure%[Standard starting position]
{%
	\includegraphics[scale=.78]{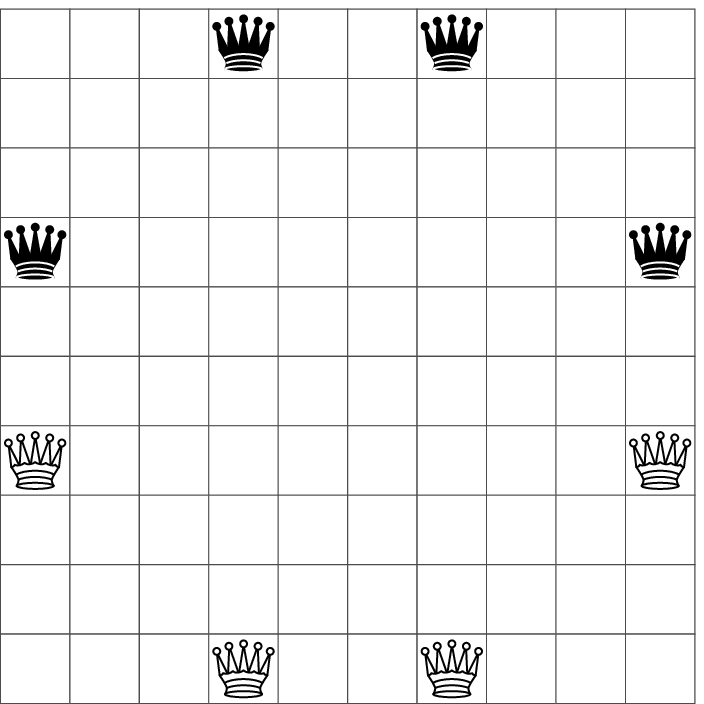}%
	\label{fig:start}%
}\hfill
\subfigure%[An endgame position]
{%
	\includegraphics[scale=.78]{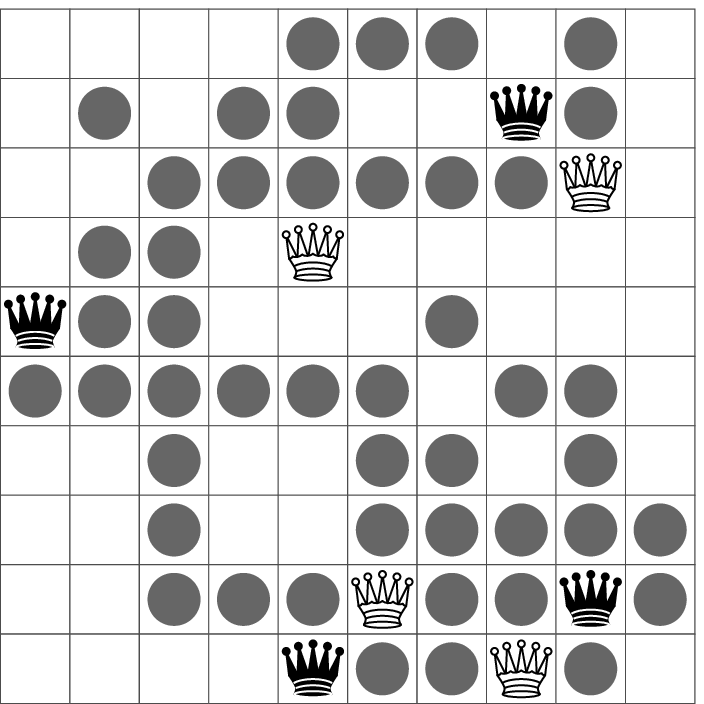}%
	\label{fig:endgame}%
}%
\hspace{.6 in}
\caption{\label{fig:amazons}
  Amazons start position and typical endgame position.}
\end{figure}

Amazons is played on a $10\times 10$ board. The standard starting position, and a typical endgame position, are shown in \fig{amazons}. Each player has four \emph{amazons}, which are immortal chess queens. White plays first, and play alternates. On each turn a player must first move an amazon, like a chess queen, and then fire an \emph{arrow} from that amazon. The arrow also moves like a chess queen. The square that the arrow lands on is burned off the board; no amazon or arrow may move onto or across a burned square. 
%(In actual play go stones are commonly used to mark the burned squares.)
There is no capturing. The first player who cannot move loses. 

Amazons is a game of mobility and control, like Chess, and of territory, like Go. The strategy involves constraining the mobility of the opponent's amazons, and attempting to secure large isolated areas for one's own amazons. In the endgame shown in \fig{amazons}, 
%each side has sealed off areas that only its amazons may reach. 
Black has access to 23 spaces, and with proper play can make 23 moves; White can also make 23 moves. 
%(Each move burns one space.) 
Thus from this position, the player to move will lose.

\section{PSPACE-completeness}

\paragraph{Formula Game.}
Schaefer \cite{Schaefer-1978-games} showed that  deciding the winner of the following two-person game is PSPACE-complete:
Let \textsf{A} be a positive CNF formula (i.e., a propositional formula in conjunctive normal form in which no negated variables occur). Each player on his move chooses a variable occurring in \textsf{A} which has not yet been chosen. After all variables have been chosen, player one wins iff \textsf{A} is true when all variables chosen by player one are set to true and those chosen by player two are set to false.

We will refer to this game as the \emph{formula game}. Our reduction consists of constructing an Amazons configuration which forces the two players to effectively play a given formula game. 

Given a positive CNF formula \textsf{A}, we build logic and wiring gadgets corresponding to the variables and the formula.
%The gadgets use a kind of ``backwards logic'': signals propagate by moving amazons backward, or by moving ``holes'' forward. 
If White plays first in a variable, a signal is enabled to flow out from it by moving White amazons backward; if Black plays first, that signal is blocked. By splitting the signals, allowing them to cross, and feeding them into a network of logic gates, eventually we may construct a particular signal line that may activate only if \textsf{A} is true under the selected variable assignment.
%; we arrange for White to win iff he activates this line.
White gains access to a large space with enough available moves for him to win only if he can activate that output signal. Black has an additional amazon isolated in a room with enough moves available to win unless White can reach the larger space.

\paragraph{Basic Wiring.}

\begin{figure}
\centering
%\hspace{1 in}
%\subfigure[Wire, turn]
%{%
%	\includegraphics[scale= .9]{turn.pdf}%
%	\label{fig:turn}%
%}\hfill
\subfigure[Wire, parity, flow limiter]
{%
	\includegraphics[scale=.78]{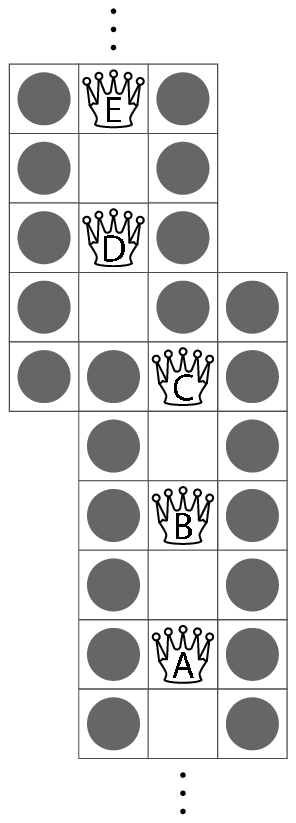}%
	\label{fig:parity}%
}\hfill
\subfigure[Turn, one way]
{%
	\includegraphics[scale=.78]{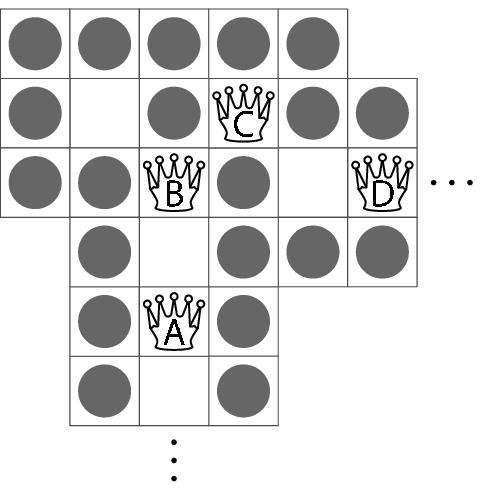}%
	\label{fig:oneway}%
}\hfill
\subfigure[Split]
{%
	\includegraphics[scale=.78]{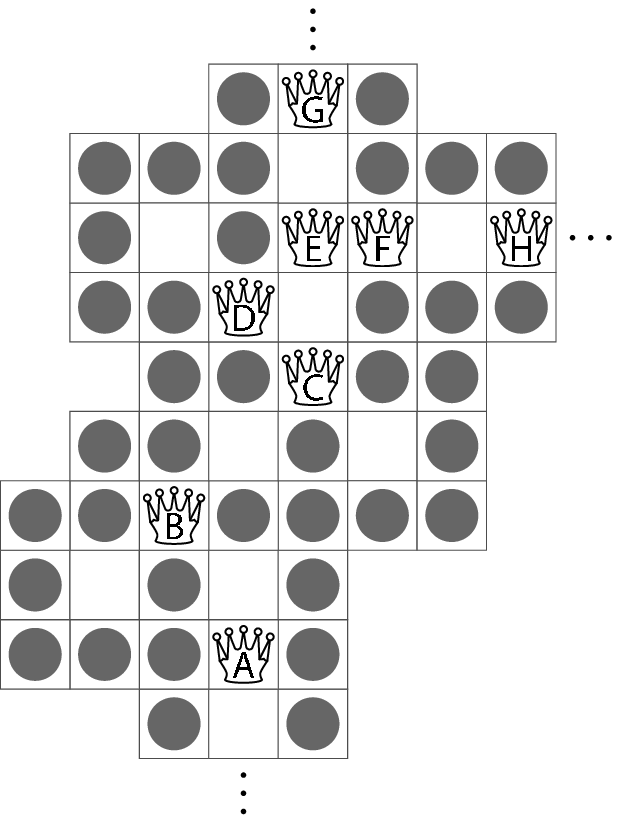}%
	\label{fig:split}%
}%
%\hspace{1 in}
\caption{\label{fig:wiring}
  Wiring gadgets.}
\end{figure}

Signals propagate along \emph{wires}. \fig{parity} shows the construction of a wire. Suppose that amazon \textsf{A} is able to move down one square and shoot down. This enables amazon \textsf{B} to likewise move down one and shoot down; \textsf{C} may now do the same. This is the basic method of signal propagation. 
%Amazons back up and shoot in the direction they moved, to free up space for the signal to propagate further. 
When an amazon moves backward (in the direction of input, away from the direction of output) and shoots backward, we will say that it has \emph{retreated}.

\newpage
\fig{parity} illustrates two additional useful features. After \textsf{C} retreats, \textsf{D} may retreat, freeing up \textsf{E}. The result is that the position of the wire has been shifted by one in the horizontal direction. Also, 
%note that 
no matter how much space is freed up feeding into the wire, \textsf{D} and \textsf{E} may still only retreat one square, because \textsf{D} is forced to shoot into the space vacated by \textsf{C}.
% Of course, \textsf{E} may move down at any time, but if it is forced to shoot up, nothing useful will be accomplished; it will not have successfully retreated.

\fig{oneway} shows how to turn corners. Suppose \textsf{A}, then \textsf{B} may retreat. Then \textsf{C} may retreat, shooting up and left; \textsf{D} may then retreat. This gadget also has another useful property: signals may only flow through it in one direction. Suppose \textsf{D} has moved and shot right. \textsf{C} may then move down and right, and shoot right. \textsf{B} may then move up and right, but it can only shoot into the square it just vacated. Thus, \textsf{A} is not able to move up and shoot up.

%\fig{oneway} shows how to turn corners. Suppose \textsf{A}, then \textsf{B} may retreat. Then \textsf{C} may retreat, shooting up and left. \textsf{D} may then retreat, freeing \textsf{E}. This gadget also has another useful property: signals may only flow through it in one direction. Suppose \textsf{E} and \textsf{D} have moved and shot right. \textsf{C} may then move down and right, and shoot right. \textsf{B} may then move up and right, but it can only shoot into the square it just vacated. Thus, \textsf{A} is not able to move up and shoot up.

By combining the horizontal parity-shifting in \fig{parity} 
with turns, we may direct a signal anywhere we wish. Using the unidirectional and flow-limiting properties of these gadgets, we can ensure that signals may never back up into outputs, and that inputs may never retreat more than a single space.

Splitting a signal is a bit trickier. The \emph{split} gadget shown in \fig{split} accomplishes this. 
\textsf{A} is the input; \textsf{G} and \textsf{H} are the outputs.  

\begin{lem}
\label{lem:split}
In a split gadget, amazons \textsf{G} and \textsf{H} may retreat iff \textsf{A} may retreat.
\end{lem}

\begin{pf}
First, observe that until \textsf{A} retreats, there are no useful moves to be made. \textsf{C}, \textsf{D}, and \textsf{F} may not move without shooting back into the square they left. \textsf{A}, \textsf{B}, and \textsf{E} may move one unit and shoot two, but nothing is accomplished by this. But if \textsf{A} retreats, then the following sequence is enabled: \textsf{B} down and right,  shoot down; \textsf{C} down and left two, shoot down and left; \textsf{D} up and left, shoot down and right three; \textsf{E} down two, shoot down and left; \textsf{F} down and left, shoot left. This frees up space for \textsf{G} and \textsf{H} to retreat.
\end{pf}

%We defer discussion of how to enable wires to cross each other until after development of the logic gadgets.

%\subsection*{Logic}
\paragraph{Logic.}

\begin{figure}
\centering
\hspace{.5 in}
\subfigure[Variable]
{%
	\includegraphics[scale=.78]{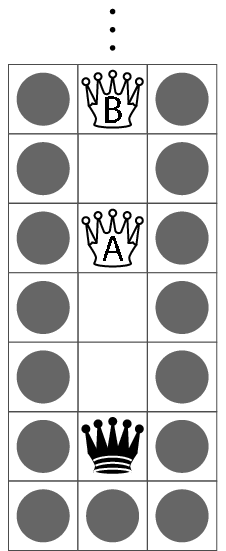}
	\label{fig:variable}%
}\hfill
\subfigure[\textsc{And}]
{%
	\includegraphics[scale=.78]{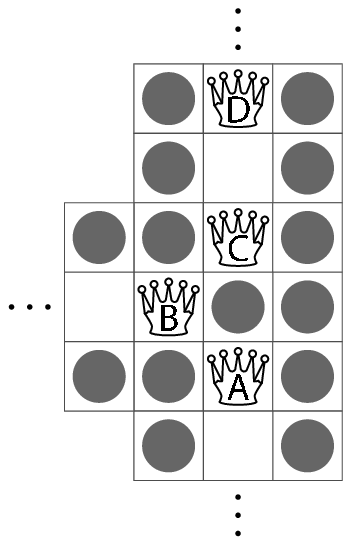}%
	\label{fig:and}%
}\hfill
\subfigure[\textsc{Or}]
%\subfigure[\textsc{Or}, choice]
{%
	\includegraphics[scale=.78]{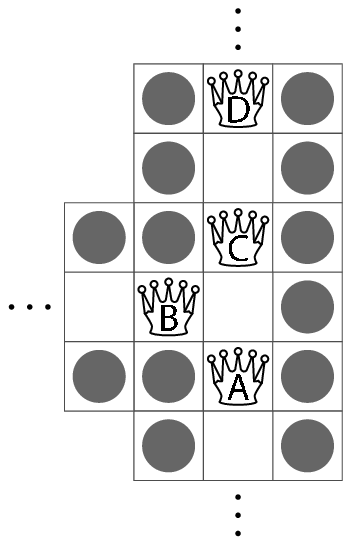}%
	\label{fig:or}%
}%
\hspace{.5 in}
\caption{\label{fig:logic}
  Logic gadgets.}
\end{figure}

The \emph{variable} gadget is shown in \fig{variable}. If White moves first in a variable, he can move \textsf{A} down, and shoot down, allowing \textsf{B} to later retreat. If Black moves first, he can move up and shoot up, preventing \textsf{B} from ever retreating. 
%Unlike the formula game, the players are not strictly forced to alternate choosing variables until none are left; later, we will show that they have no better strategy.

The \textsc{And} and \textsc{Or} gadgets are shown in \figs~\ref{fig:and} and \ref{fig:or}. In each, \textsf{A} and \textsf{B} are the inputs, and \textsf{D} is the output. Note that we assume the inputs are protected with flow limiters (\fig{parity}), so that no input may retreat more than one square (otherwise the \textsc{And} might incorrectly activate).

\begin{lem}
\label{lem:and}
In an \textsc{And} gadget, amazon \textsf{D} may retreat iff \textsf{A} and \textsf{B} first retreat.
\end{lem}

\begin{pf}
No amazon may usefully move until at least one input retreats. If \textsf{B} retreats, then a space is opened up, but \textsf{C} is unable to retreat there; similarly if just \textsf{A} retreats. But if both inputs retreat, then \textsf{C} may move down and left, and shoot down and right, allowing \textsf{D} to retreat.
\end{pf}

\begin{lem}
\label{lem:or}
In an \textsc{Or} gadget, amazon \textsf{D} may retreat iff either \textsf{A} or \textsf{B} first retreats.
\end{lem}

\begin{pf}
Similar to above.
\end{pf}

%The \textsc{And} and \textsc{Or} gadgets are shown in \figs~\ref{fig:and} and \ref{fig:or}. In each, \textsf{A} and \textsf{B} are the inputs, and \textsf{F} is the output. Note that we assume the inputs are protected with flow limiters (\fig{parity}), so that no input may retreat more than one square (otherwise the \textsc{And} might incorrectly activate).

%\begin{lem}
%\label{lem:and}
%In an \textsc{And} gadget, amazon \textsf{F} may retreat iff \textsf{A} and \textsf{B} first retreat.
%\end{lem}

%\begin{pf}
%No piece may usefully move until at least one input retreats. If \textsf{B} and \textsf{D} retreat, then \textsf{D}'s space is opened up, but \textsf{E} is unable to retreat there; similarly if just \textsf{A} and \textsf{C} retreat. But if both inputs retreat, then \textsf{E} may move down and left, and shoot down and right, allowing \textsf{F} to retreat.
%\end{pf}

%\begin{lem}
%\label{lem:or}
%In an \textsc{Or} gadget, amazon \textsf{F} may retreat iff either \textsf{A} or \textsf{B} first retreats.
%\end{lem}

%\begin{pf}
%Similar to above.
%\end{pf}

%The \textsc{Or} gadget has another interesting property. If we view \textsf{B} as an output instead of an input, then this becomes a \emph{choice} gadget: if \textsf{A} retreats then either \textsf{F} or \textsf{B} may retreat, but not both. We will use this property for an auxiliary result; we don't use choice gadgets in the main construction.

\paragraph{Crossover.}

\begin{figure}
\centering
\hspace{1 in}
\subfigure[Half crossover]
{%
	\includegraphics[scale=.78]{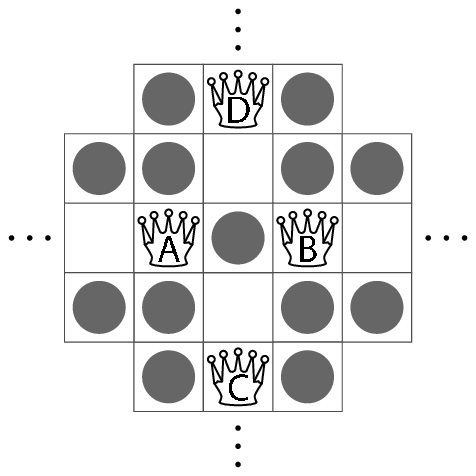}%
	\label{fig:half-crossover}%
}\hfill
\subfigure[Crossover]
{%
	\includegraphics[scale=.52]{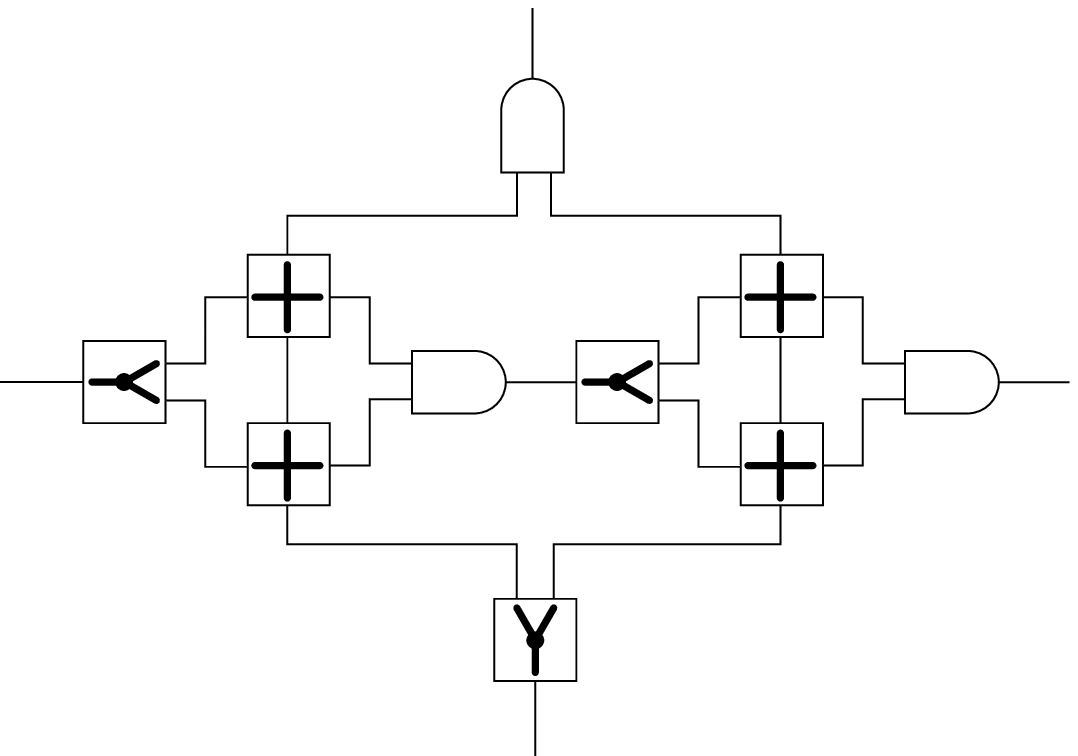}%
	\label{fig:crossover}%
}%\hfill
%\subfigure[Crossover]
%{%
%	\includegraphics[scale=.7]{crossoverclean.eps}%
%	\label{fig:crossover}%
%}%\hfill
\hspace{1 in}
\caption{\label{fig:crossovers}
  Crossover gadgets.}
\end{figure}

We develop the ability to cross wires over each other in two steps. The first step is the \emph{half-crossover} gadget, shown in \fig{half-crossover}. Using half crossovers, splits, and \textsc{And}s, we can make a full \emph{crossover} gadget, shown schematically in \fig{crossover}. (Splits are shown with a forking symbol, half crossovers with a plus symbol, and \textsc{And}s with a standard digital logic symbol.)

We note that this crossover construction is essentially the same as that used in Hearn and Demaine's development of nondeterministic constraint logic \cite{NCL_TCS}; the only differences are that in \cite{NCL_TCS}, the gates are reversible, and the half crossover is also made of primitive gates. 
%(We could similarly make the half crossover out of gates here, but the explicit construction is simpler.) This crossover construction would seem to have a wide range of applicability for reductions of this type, but does not seem to be widely known.

\begin{lem}
\label{lem:half-crossover}
In a half-crossover gadget, the number of output amazons \textsf{B}, \textsf{D} that may retreat equals the number of input amazons \textsf{A}, \textsf{C} that have retreated (i.e., 0, 1, or 2); any such retreats are possible.
\end{lem}

\begin{pf}
If no inputs have retreated, then there are no moves allowing an output to retreat.

If \textsf{A} retreats, then either \textsf{B} or \textsf{D}, but not both, may move to one space below \textsf{D}, and shoot down and left. (\textsf{C} could instead move inward, but this would not be useful.) If \textsf{C} retreats, then similar reasoning applies.

If both inputs retreat, then there is space for both outputs to retreat.
\end{pf}

%
%\begin{lem}
%\label{lem:half-crossover}
%In a half-crossover gadget, the number of output amazons \textsf{D}, \textsf{H} that may retreat equals the number of input amazons \textsf{A}, \textsf{E} that have retreated (i.e., 0, 1, or 2); any such retreats are possible.
%\end{lem}

%\begin{pf}
%If no inputs have retreated, then there are no moves allowing an output to retreat.

%If \textsf{A} and \textsf{B} retreat, then either \textsf{C} or \textsf{G} may move to one space below \textsf{G}, and shoot down and left. This will enable either \textsf{D} or \textsf{H}, but not both, to retreat. (\textsf{E} and \textsf{F} could instead move inward, but this would not be useful.) If \textsf{E} and \textsf{F} retreat, then similar reasoning applies.

%If both inputs retreat, then there is space for both outputs to retreat.
%\end{pf}

\begin{lem}
\label{lem:crossover}
In a crossover gadget, the right \textsc{And}'s output amazon may retreat iff the left split's input amazon retreats; the top \textsc{And}'s output amazon may retreat iff the bottom split's input amazon retreats.
\end{lem}

\begin{pf}
First, it is clear that if one input retreats, the corresponding output may also retreat; simply choose the straight-through path to activate for each half crossover. If both inputs retreat, activating both half-crossover outputs allows both crossover outputs to retreat.

Suppose the left split's input has not retreated. Then at most one input to the left \textsc{And} may retreat, because the bottom-left half crossover can have at most one input, and thus output, retreat (by Lemma \ref{lem:half-crossover}). Therefore, the left \textsc{And}'s output amazon may not retreat. By the same reasoning, the right \textsc{And}'s output amazon may not retreat either.
A similar argument shows that if the bottom split's input has not retreated, the top \textsc{And}'s output may not retreat.
\end{pf}

\paragraph{Winning.}

\begin{figure}
\centering
\includegraphics[scale=.78]{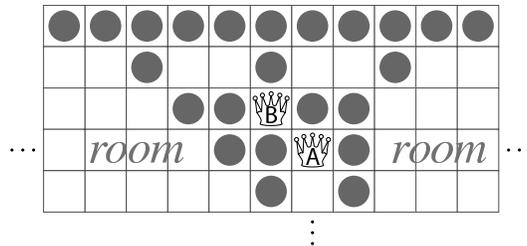}
\caption{\label{fig:victory}
Victory gadget.%
}
\end{figure}

We will have an \textsc{And} gadget whose output may be activated only if the formula is true under the chosen assignment. We feed this signal into a \emph{victory} gadget, shown in \fig{victory}.  There are two large rooms available. The sizes are equal, and such that if White can claim both of them, he will win, but if he can claim only one of them, then Black will win. 

\begin{lem}
\label{lem:victory}
In a victory gadget, White may reach both of the rooms iff \textsf{A} may retreat. Otherwise, he may reach either one, but not both, of them.
\end{lem}

\begin{pf}
If \textsf{B} moves before \textsf{A} has retreated, then it must shoot so as to block access to one room or the other; it may then enter and claim the accessible room. If \textsf{A} first retreats, then \textsf{B} may move up and left, and shoot down and right two, leaving the way clear to enter and claim the left room, then back out and enter and claim the right room.
\end{pf}

%\begin{pf}
%If \textsf{C} moves before \textsf{A} and \textsf{B} have retreated, then it must shoot so as to block access to one room or the other; it may then enter and claim the accessible room. If \textsf{A} and \textsf{B} first retreat, then \textsf{C} may move up and left, and shoot down and right two, leaving the way clear to enter and claim the left room, then back out and enter and claim the right room.
%\end{pf}

\newpage

\begin{thm}
\label{thm:pspace}
Amazons is PSPACE-complete.
\end{thm}

\begin{pf}
Given a positive CNF formula \textsf{A}, we construct a corresponding Amazons position, as described above. The reduction may be done in polynomial time: if there are $k$ variables and $l$ clauses, then there need be no more than $(kl)^2$ crossover gadgets to connect each variable to each clause it occurs in; all other aspects of the reduction are equally obviously polynomial.

If the players alternate choosing variables, then when all variables have been chosen, White will be able to activate wires leading from only those variables he has chosen; these are just the variables assigned to true in the formula game. Since \textsf{A} contains no negated variables, 
White will thus be able eventually to reach both rooms of the victory gadget iff \textsf{A} is true under the variable assignment corresponding to the players' choices (using Lemmas \ref{lem:split}, \ref{lem:and}, \ref{lem:or}, \ref{lem:crossover}, and \ref{lem:victory}). White will then have more moves available than Black, and win; otherwise, Black's extra room will give him more moves than White, and Black will win.

Suppose a player makes a move which does not choose a variable, before all variables have been chosen. This can have no effect on the other player, apart from allowing him to choose two variables in a row, because the Black and White amazons may only interact within variable gadgets. A player who chooses two variables in a row may finish with at least the same set of variables chosen as he would otherwise. Therefore, not playing in accordance with the formula game does not allow a player to win if he could not otherwise win.

Therefore, a player may win the Amazons game iff he may win the corresponding formula game, and Amazons is PSPACE-hard. 

Since the game must end after a polynomial number of moves, it is possible to perform a search of all possible move sequences using polynomial space, thus determining the winner. Therefore, Amazons is in PSPACE.
%
%Note that since \textsc{A} contains no negated variables, we do not need to route any signals from 
%
%
%
%The wiring gadgets give us the freedom to route a wire from each variable gadget to each clause in which the corresponding variable is used.
\end{pf}

\section{Simple Amazons Endgames}
A \emph{simple Amazons endgame} is an Amazons position in which the Black and White amazons are completely separated by burned squares. There can thus be no interaction between the amazons, and the winner is determined by which player can make the most moves in his own territory. 
Buro \cite{Buro-2000} showed that it is NP-complete to decide whether a player may make a given number of moves from an individual territory containing only his amazons. Buro first proved  NP-completeness of the Hamilton circuit problem for cubic subgraphs of the integer grid, and then reduced from that problem.
As a result,
%Buro showed that
 deciding the outcome of a simple Amazons endgame is NP-equivalent (that is, it can be decided with a polynomial number of calls to an algorithm for an NP-complete problem, and vice versa). 
%
%Buro first proved that the Hamilton circuit problem for cubic subgraphs of the integer grid is NP-complete, and then reduced from that problem to Amazons. % deciding whether a player may make a given number of moves from a position containing only his amazons.
%
Our gadgets provide an alternate proof.
% Like Buro, we show that it is NP-complete to decide whether a player may make a given number of moves from a position containing only his amazons.

\begin{thm}
Deciding the outcome of a simple Amazons endgame is NP-equivalent.
\end{thm}

\begin{pf}
We reduce SAT to a single-color Amazons position. Given a propositional formula \textsf{A}, we construct the same position as in Theorem \ref{thm:pspace}, with the following modifications.
%
%construct a corresponding network of \textsc{And}s, \textsc{Or}s, splits, and crossovers, as in Theorem \ref{thm:pspace}, with the final \textsc{And} output leading to a victory gadget. 
%We use one variable gadget for each variable in the formula, but
We remove the Black amazons, then connect each variable output to an input of a half-crossover, and block off the second input with burned squares. White may then activate a path from each variable, but at the corresponding half-crossover, only one output Amazon may retreat. The half-crossover thus serves as a two-way choice gadget. We connect one output path to the non-negated occurrences of the corresponding variable in the formula, and the other output path to the negated occurrences.

Then, White may reach both rooms of the victory gadget iff \textsf{A} is satisfiable, by choosing the correct set of half-crossover output paths. Therefore, it is NP-hard to decide whether a player may make a given number of moves from a position containing only his amazons. We may nondeterministically guess a satisfying move sequence and verify it in polynomial time; therefore, the problem is NP-complete.
As in \cite{Buro-2000}, it follows automatically that deciding the winner of a simple Amazons endgame is NP-equivalent.
%By adjoining to this position a long isolated corridor containing a single Black amazon, we create a simple Amazons endgame. We can decide the winner of the endgame by deciding whether White can make more moves than the length of the corridor, and we 
\end{pf}

%\begin{wrapfigure}[13]{r}{1.8 in}
%  \centering
%  \includegraphics[scale=0.49]{donkey-clean-cleanclean.eps}
%  \caption{\label{sliding block puzzles}
%    The Donkey Puzzle: move the large square to the bottom center.}
%\end{wrapfigure}

%
%\paragraph*{Roadmap.}
%\sect{definitions} defines the necessary hinged dissection terminology and outlines the 
%decision questions we will be interested in.
%\sect{ncl} describes Nondeterministic Constraint Logic, the tool by which we show PSPACE-hardness.
%\sect{hardness} gives the constructions that show all of our decision questions PSPACE-hard.
%\sect{related} applies these results to re-derive the PSPACE-hardness of 2-d linkage equivalence, previously
%shown hard by \cite{tree-linkage}.
%\sect{conclusion} summarizes our results.

%

%\begin{figure}
%\centering
%\hspace{.1 in}
%\subfigure[\textsc{And} vertex]
%{%
%	\includegraphics[scale= .45]{and-cleanclean.eps}%
%	\label{fig:and}%
%}\hfill
%\subfigure[\textsc{Or} vertex]
%{%
%	\includegraphics[scale=.45]{or-clean.eps}%
%	\label{fig:or}%
%}%
%\hspace{.1 in}
%\caption{\label{fig:vertices}
%  NCL vertex gadgets.}
%\end{figure}

%\paragraph{Vertex gadgets.} \fig{vertices} shows the vertex gadgets. Each gadget is made of several pieces. The pieces
%%\paragraph{Vertex gadgets.} Figure \ref{fig:vertices} shows the vertex gadgets. Each gadget is made of several pieces. The pieces
%are joined together by hinges; each sliding connection in \fig{vertices} is shorthand for a hinged
%connection as shown in \fig{slider}.
%%

\section{Conclusion}

We have shown that generalized Amazons is PSPACE-complete, indicating that it is highly unlikely that an efficient algorithm for optimal play exists.

%\paragraph{Future Work.}
Our reduction uses a number of amazons polynomial in the input formula length. 
We speculate that Amazons remains PSPACE-complete even when only a constant number of amazons is used -- perhaps as few as one per player. However, a reduction to prove this would need a completely different approach to the problem than the one presented here.

\bibliography{combinatorialgames,complexity,motionplanning,polytopes,pushingblocks}
\bibliographystyle{plain}

%\begin{thebibliography}{10}

%\end{thebibliography}

\end{document}